\documentclass[twocolumn,showpacs]{revtex4}

\usepackage{graphicx}
\usepackage{dcolumn}
\usepackage{bm}
\usepackage{latexsym}
\usepackage{tabularx}
\usepackage{hhline}

\begin{document}


\title{An orbital-free density functional method based on inertial fields.}

\author{Kriszti\'{a}n Palot\'{a}s$^1$ and Werner A.\ Hofer$^{1,2}$}
\affiliation{
         $^1$ Surface Science Research Centre and Department of Physics,
         University of Liverpool, Liverpool L69 3BX, Britain \\
         $^2$ Donostia International Physics Center, San Sebastian,
         Spain}

\begin{abstract}
In this paper we revisit the Levy-Perdew-Sahni equation. We
establish that the relation implicitly contains the conservation of
energy density at every point of the system. The separate
contributions to the total energy density are described in detail,
and it is shown that the key difference to standard density
functional methods is the existence of a general
exchange-correlation potential, which does not explicitly depend on
electron charge. We derive solutions for the hydrogen-like atoms and
analyse local properties. It is found that these systems are stable
due to the existence of a vector potential ${\bf A}$, related to
electron motion, which leads to two general effects: (i) The root of
the charge density acquires an additional complex phase; and (ii)
for single electrons, the vector potential cancels the effect of
electrostatic repulsions. We determine the density of states of a
free electron gas based on this model and find that the
vectorpotential also accounts for the Pauli exclusion principle.
Implications of these results for direct methods in density
functional theory are discussed. It seems that the omission of
vector potentials in formulating the kinetic energy density
functionals may be the main reason that direct methods so far are
not generally applicable. Finally, we provide an orbital free
self-consistent formulation for determining the groundstate charge
density in a local density approximation.
\end{abstract}


\maketitle

\section{Introduction}

A key innovation in theoretical solid state physics in the last
fifty years was the reformulation of quantum mechanics in a density
formalism, based on the Hohenberg-Kohn theorem \cite{hk64}. Despite
initial resistance, in particular from quantum chemists, the method
has replaced previous frameworks and provides, to date, the most
advanced theoretical model for the calculation of atoms, solids, and
liquids. However, its implementation relies on a rather cumbersome
detour. While the Hohenberg-Kohn theorem is formulated exclusively
in terms of electron densities and energy functionals, calculations
today are based almost exclusively on the specifications given by
Kohn and Sham one year after the initial theorem was made public
\cite{ks65}. While this procedure is generally successful, and
implemented today in numerical methods optimized for efficiency (see
e.g the ingenious way ionic and electron degrees of freedom are
treated on much the same footing following a method developed by Car
and Parinello \cite{cp85}), it is highly inefficient in one crucial
conceptual point: If, according to the Hohenberg-Kohn theorem, the
electron density is the only physically relevant variable of the
system, then solving the Schr\"odinger equation, setting up the
eigenvectors, and computing the density of electrons is an
operation, which creates a vast amount of redundant information.
Every information, pertaining to the solution of the single-particle
Schr\"odinger equation and the summation of single electron charges
is discarded at the end of every step in the iteration cycle. One
could therefore say that more than 90\% of the information created
in today's simulations is actually irrelevant. The question thus
arises: {\em Do we have to create this information at all, or can we
find a more direct way to arrive at the groundstate density of
electrons without this cumbersome detour via the single-particle
Schr\"odinger equation?}

In 1984, Levy, Perdew, and Sahni (LPS) published an intriguing
relation \cite{levy84}, in which the density of charge is described
by a second order differential equation without any reference to
Kohn-Sham orbitals \cite{ks65}. It introduced the possibility to
formulate the many-electron problem with just one quantity, the
electron density. In this respect it seemed to be more in line with
the original Hohenberg-Kohn theorem \cite{hk64} than standard methods
in density functional theory, based on the Kohn-Sham equations
\cite{ks65}. In the last twenty years, developments in this field
focussed on a search for suitable density functional approximations,
especially for kinetic energy functionals. It is by now an expansive
field of research in theoretical solid state physics and quantum
chemistry. The main results of subsequent work on the LPS relation
\cite{levy84} are: the usage of systematically constructed Harriman
orbitals \cite{harriman81} in a three-dimensional generalisation
\cite{ghosh84}; the introduction and analysis of the Pauli potential
\cite{march85,levy88}; the analysis of the uniqueness and asymptotic
behaviour of the local kinetic energy \cite{yang96}; the properties
of the kinetic energy density \cite{sim03}; and weighted or averaged
density approximation \cite{chacon85,alonso78,garcia96}. For a
comprehensive review, see \cite{wangcarter}. The list
is, of course, not complete. In particular it does not contain
references to the development of the Kohn-Sham theory, e.g. the
continued improvement of exchange-correlation functionals, as this
is not the topic of the present paper. Here, we want to reexamine
the original LPS relation and its properties.

The outline of the paper is as follows: in Section \ref{loc_prop} we
describe the local properties of the LPS equation and find that it
entails conservation of energy density throughout a quantum
mechanical system. In Section \ref{atom_ana} we give the results of
analytic applications of the LPS relation for the hydrogen atom. We
find that the effective potential is the sum of a non-zero Hartree
potential and an equally non-zero general exchange-correlation
potential. The effective potential is zero due to cancellation of
the two contributions. In Section \ref{gen_xc_pot} we analyse the
physical origin of this cancellation and find that motion of single
electrons creates a vector potential ${\bf A}$, called the {\em
inertial field} due to its relation to electron motion. Due to this
potential the root of the charge density acquires an additional
complex phase. The phase-shift is in line with an Aharonov-Bohm
effect \cite{aharonov}. Based on these findings we formulate the general
problem for an $N$-electron system in Section \ref{gen_prob}.
Finally, in Section \ref{sum_dis} we discuss
the results and estimate their importance for the development of
orbital free density functional methods.

\section{Local properties}\label{loc_prop}

A closer analysis reveals that the genuine novelty of the LPS
relation seems to have been disregarded to this date. It is the
conservation of energy at a local level. This is well in advance of
the single-particle Schr\"odinger equation or a many-body framework,
where energy is conserved only globally.

\subsection{General relations}

We start with the LPS relation,
\begin{equation}
\label{lps_phi} \left[-\frac{1}{2}{\nabla}^2+v_{ext}({\bf
r})+v_{eff}({\bf r})\right] \rho({\bf r})^{1/2}
= \mu \rho({\bf r})^{1/2},
\end{equation}
rearranging as
\begin{equation}
-\frac{1}{2}\frac{\nabla^2\rho({\bf r})^{1/2}}{\rho({\bf r})^{1/2}}+
v_{ext}({\bf r})+v_{eff}({\bf r})=\mu,
\end{equation}
and multiplying by $\rho({\bf r})$ results in
\begin{equation}
\label{locenergy} -\frac{1}{2}\rho({\bf r})^{1/2}\nabla^2\rho({\bf
r})^{1/2}+ v_{ext}({\bf r})\rho({\bf r})+v_{eff}({\bf r})\rho({\bf
r})= \mu\rho({\bf r}).
\end{equation}
We will show that the potential at a point ${\bf r}$ multiplied by
the charge density at this point, describes the potential energy
density. Thus the equation is nothing else but a description of
energy conservation for every point of the system,
\begin{equation}
\label{locenergy2} t({\bf r})+\varepsilon_{ext}({\bf
r})+\varepsilon_{eff}({\bf r})= \varepsilon_{tot}({\bf r}).
\end{equation}
Here, we have symbolized the term $\mu\rho({\bf r})$ by a total
energy density $\varepsilon_{tot}({\bf r})$. Each term
refers to a corresponding energy density. The resulting equation
is equivalent to the LPS relation, which therefore contains energy
conservation also in its general form. Since the relation between
potentials and energy densities may not be directly accessible, we
derive them in the following from fundamental considerations.

The kinetic energy density $t({\bf r})$ is formulated for
interacting bosonic \cite{levy84} particles. It can exactly be
rewritten in a more useful manner,
\begin{equation}
\label{eq:t} t({\bf r})=-\frac{1}{4}\nabla^2\rho({\bf r})+
\frac{1}{8}\frac{[\nabla\rho({\bf r})]^2}{\rho({\bf r})}=
t(\rho({\bf r}),\nabla\rho({\bf r}),\nabla^2\rho({\bf r})).
\end{equation}
The usefulness of this formulation will be seen further down. It
should be noted that the second term of the above expression is
the von Weizs\"acker kinetic energy density \cite{weiz}. For
electron charge contained in an infinitesimal volume $dV$ around a
point ${\bf r}$, the contribution to the interaction energy
between electrons and nuclei $E_{ext}({\bf r})$ will be
\begin{equation}
dE_{ext}({\bf r})=-\sum_{i=1}^M\frac{Z_i}{|{\bf r}-{\bf R}_i|}
\rho({\bf r})dV,
\end{equation}
where $M$ is the number of nuclei and $Z_i$ the charge of the
$i$th nucleus. The electron-nuclear energy density at a point
${\bf r}$ is then
\begin{equation}
\label{eq:eext} \varepsilon_{ext}({\bf r})=\frac{dE_{ext}({\bf
r})}{dV}= -\sum_{i=1}^M\frac{Z_i\rho({\bf r})}{|{\bf r}-{\bf R}_i|}=
v_{ext}({\bf r})\rho({\bf r}).
\end{equation}
The effective potential contains the explicit electron-electron
interaction (Hartree potential, $v_H({\bf r})$) and a generalized
exchange-correlation potential (gxc-potential, $v_{gxc}({\bf r})$)
which includes the exchange and correlation effects, as well as the
potential contributions due to Pauli exclusion,
$v_{eff}({\bf r})=v_H({\bf r})+v_{gxc}({\bf r})$. The contribution
to the Hartree energy from the electron distribution between two
infinitesimal volume $dV$ and $dV'$ around points ${\bf r}$ and
${\bf r'}$ respectively is
\begin{equation}
d^2E_H({\bf r},{\bf r'})= \frac{\rho({\bf r})dV\rho({\bf
r'})dV'}{|{\bf r}-{\bf r'}|}.
\end{equation}
Integrating over $dV'$ and rearranging the results leads to the
Hartree energy density at point ${\bf r}$,
\begin{equation}
\label{eq:eh} \varepsilon_H({\bf r})=\frac{dE_H({\bf r})}{dV}=
\int d^3r'\frac{\rho({\bf r'})\rho({\bf r})}{|{\bf r}-{\bf r'}|}=
v_H({\bf r})\rho({\bf r}).
\end{equation}
The same behaviour is found for the general exchange-correlation
energy density,
\begin{equation}
\label{eq:exc} \varepsilon_{gxc}({\bf r})=\frac{dE_{gxc}({\bf
r})}{dV}= v_{gxc}({\bf r})\rho({\bf r}),
\end{equation}
although at this point the exact form of $v_{gxc}({\bf r})$ is not
known.

So far we have proved that the left hand sides of
Eqs.\ (\ref{locenergy}) and (\ref{locenergy2}) are the same term by
term. In order to prove that the total energy density is equal to
the chemical potential times the electron density
($\varepsilon_{tot}({\bf r})=\mu\rho({\bf r})$) we employ the
variational principle to minimize the total energy. This procedure
also allows to determine the relationship between the functional
derivatives of the energy terms, and the corresponding energy
densities. Integrating Eq.\ (\ref{locenergy2}) over the whole space
the total energy will be
\begin{eqnarray}
E_{tot}[\rho]&=&T[\rho]+E_{ext}[\rho]+E_{H}[\rho]+E_{gxc}[\rho]\nonumber\\
&=&\int d^3r\left[-\frac{1}{4}\nabla^2\rho({\bf r})+
\frac{1}{8}\frac{[\nabla\rho({\bf r})]^2}{\rho({\bf r})}\right]
\nonumber\\&&-\int d^3r\sum_{i=1}^M\frac{Z_i\rho({\bf r})}{|{\bf
r}-{\bf R}_i|} +\frac{1}{2}\int \int d^3rd^3r' \frac{\rho({\bf
r})\rho({\bf r}')}{|{\bf r}-{\bf r}'|}\nonumber\\ &&+\int d^3r
v_{gxc}({\bf r})\rho({\bf r}).
\end{eqnarray}
The variational principle, including the condition for the total
number of electrons $N$ with a Lagrange multiplier $\mu$, provides
us with the groundstate energy,
\begin{equation}
\frac{\delta}{\delta\rho}\left[E_{tot}[\rho]- \mu\left(\int
d^3r\rho({\bf r})-N\right)\right]=0.
\end{equation}
This leads to the following result:
\begin{eqnarray}
\label{fderE}
\frac{\delta E_{tot}}{\delta\rho}&=&\frac{\delta T}{\delta\rho}
+\frac{\delta E_{ext}}{\delta\rho}+\frac{\delta E_H}{\delta\rho}
+\frac{\delta E_{gxc}}{\delta\rho}\\
&=&-\frac{1}{4}\frac{\nabla^2\rho({\bf r})}{\rho({\bf r})}+
\frac{1}{8}\frac{[\nabla\rho({\bf r})]^2}{\rho({\bf r})^2}-
\sum_{i=1}^M\frac{Z_i}{|{\bf r}-{\bf R}_i|}\nonumber\\
&&+\int d^3r'\frac{\rho({\bf r}')}{|{\bf r}-{\bf r}'|}+ v_{gxc}({\bf
r})+\rho({\bf r})\frac{dv_{gxc}({\bf r})}{d\rho({\bf r})}
=\mu.\nonumber
\end{eqnarray}
It must be noted that the energy term $\mu$ carries two separate
meanings: mathematically, it is a Lagrange multiplier; physically,
it is also the highest occupied level of the groundstate solution of
the Kohn-Sham equations as well as the negative of the ionization
energy in the exact DFT as was proved earlier by Perdew et al.
\cite{perdew82}. In the LPS relation, it describes the eigenvalue of
the problem, and the change of the total energy, if one electron is
removed from the system if $N>>1$ \cite{levy84}.

Here, we find the first unconventional implication of the LPS
equation. Since it is not specified, in the derivation of the
chemical potential from the LPS relation \cite{levy84}, which
electron is actually removed from the system, the chemical potential
must be equal for the removal of any single electron. In this case,
it is only possible, like in standard DFT, to think of a
many-electron system as composed of a discrete number of electrons
with different energy levels, if we assume that the system will
balance the removal of one electron by a change of all electronic
eigenstates so that the highest electronic eigenstate remains empty.
But from the previous derivation we may also conclude that $E_{tot}$
describes the total energy of the system. In this case, $\mu$ must
have a double meaning: it is, firstly, the energy eigenvalue of the
$N$-electron system; and $\mu N$ is equal to the total energy.
Secondly, it is also the chemical potential of the system, or the
negative ionization energy. This result differs from standard DFT,
where the total energy is a sum containing the discrete energy
levels of every single eigenstate.

To understand the difference we revert to the proof by LPS that
$\mu$ is the negative ionization energy of a single electron removed
from the $N$-electron system, if the number of electrons is
sufficiently large \cite{levy84}. Suppose now, we remove one
electron from the $N$-electron system. The related chemical
potential is symbolized by $\mu_N$, the negative ionization energy
of one electron of the $N$-electron system. Now, the system contains
$N-1$ electrons. This means that the effective potential will be
lower than in the first case:
\begin{equation}
v_{eff,N-1} < v_{eff,N}.\nonumber
\end{equation}
Considering local energy conservation this implies also that
$\mu_{N-1}<\mu_N$.

For calculating $\delta T/\delta\rho$ we used
Eq.\ (\ref{eq:t}) and the following rule of functional derivatives:
\begin{eqnarray}
\text{if}&\;& T[\rho]=\int d^3rt(\rho({\bf r}),\nabla\rho({\bf
r}),\nabla^2\rho({\bf r})), \nonumber\\ \text{then}&\;&
\frac{\delta T}{\delta\rho}=\frac{\partial t}{\partial\rho}
-\nabla\frac{\partial t}{\partial\nabla\rho}
+\nabla^2\frac{\partial t}{\partial\nabla^2\rho}.
\end{eqnarray}
Multiplying Eq.\ (\ref{fderE}) by the electron density $\rho({\bf r})$
gives,
\begin{eqnarray}
\frac{\delta E_{tot}}{\delta\rho}\rho&=&\frac{\delta
T}{\delta\rho}\rho +\frac{\delta
E_{ext}}{\delta\rho}\rho+\frac{\delta E_H}{\delta\rho}\rho
+\frac{\delta E_{gxc}}{\delta\rho}\rho\nonumber\\
&=&-\frac{1}{4}\nabla^2\rho({\bf r})+
\frac{1}{8}\frac{[\nabla\rho({\bf r})]^2}{\rho({\bf r})}-
\sum_{i=1}^M\frac{Z_i}{|{\bf r}-{\bf R}_i|}\rho({\bf r})\nonumber\\
&&+\int d^3r'\frac{\rho({\bf r}')\rho({\bf r})}{|{\bf r}-{\bf
r}'|}+ v_{gxc}({\bf r})\rho({\bf r})\nonumber\\
&&+\rho({\bf r})^2\frac{dv_{gxc}({\bf r})}{d\rho({\bf r})}
=\mu\rho({\bf r}).
\end{eqnarray}
Comparing this with
Eqs.\ (\ref{locenergy}),(\ref{locenergy2}),(\ref{eq:t}),(\ref{eq:eext}),(\ref{eq:eh})
and (\ref{eq:exc}) it can be concluded that the energy densities for
the different terms are,
\begin{eqnarray}
\varepsilon_{tot}({\bf r})&=&\frac{\delta E_{tot}}{\delta\rho}
\rho({\bf r})=\mu\rho({\bf r}),\\
t({\bf r})&=&\frac{\delta T}{\delta\rho}\rho({\bf r})=
-\frac{1}{4}\nabla^2\rho({\bf r})+\frac{1}{8}\frac{[\nabla\rho({\bf r})]^2}{\rho({\bf r})},\\
\varepsilon_{ext}({\bf r})&=&\frac{\delta E_{ext}}{\delta\rho}
\rho({\bf r})=v_{ext}({\bf r})\rho({\bf r}),\\
\varepsilon_H({\bf r})&=&\frac{\delta E_H}{\delta\rho}
\rho({\bf r})=v_H({\bf r})\rho({\bf r}),\\
\label{exc} \varepsilon_{gxc}({\bf r})&=&\frac{\delta
E_{xc}}{\delta\rho} \rho({\bf r})=v_{gxc}({\bf r})\rho({\bf
r})\;,\;\frac{dv_{gxc}}{d\rho}=0.
\end{eqnarray}
Here, we arrive at an interesting consequence for the general
exchange-correlation potential. The energy density for each separate
term is described by the functional derivative times the electron
density, even for the most complicated kinetic energy term. If this
procedure holds also for the general exchange-correlation term, as
written in Eq.\ (\ref{exc}), then the general exchange-correlation
potential does not explicitly depend on the electron density. This
is in marked contrast to the standard formulations in DFT, where
exchange-correlations not only depend on the density of charge, but
are generally parametrized in terms of the density (e.g.
\cite{wigner34}).

The energy conservation on a local level, i.e., at every point ${\bf
r}$ of the system, can thus be directly deduced from the LPS
relation. Comparing the energy densities with the functional
derivatives of the corresponding energy terms has another
consequence for the kinetic energy density $t({\bf r})$ and the
kinetic energy functional $T[\rho]=\int d^3rt({\bf r})$. Starting
from Eq.\ (\ref{eq:t}), integrating over the whole space, performing
the functional derivative, and multiplying by the electron density
leads again to the kinetic energy density. From that cycle it can be
clearly seen that the kinetic energy density {\em cannot} contain
arbitrary terms whose space integral vanishes. This finding is in
contrast to previous assumptions \cite{wang92,yang96,sim03}. It is
justified by the local energy conservation of the LPS relation
itself. Let us analyse the asymptotic behaviour of Eq.\ (\ref{fderE})
as $|{\bf r}|\rightarrow\infty$. After considering that $\lim_{|{\bf
r}|\rightarrow\infty}v_{ext}({\bf r})=0$ and $\lim_{|{\bf
r}|\rightarrow\infty}v_{eff}({\bf r})=0$ \cite{levy84} we arrive at
\begin{equation}
\lim_{|{\bf r}|\rightarrow\infty}\frac{\delta T}{\delta\rho}=
\lim_{|{\bf r}|\rightarrow\infty}\frac{t({\bf r})}{\rho({\bf r})}=\mu
\end{equation}
which was found by Yang et al.\ \cite{yang96}.
This means that there is no need of approximating the kinetic energy
functional, it is exact and unique and has the correct asymptotic
behaviour, although it describes non-interacting bosons.

The detailed analysis reveals three important properties of the LPS
equation:
\begin{itemize}
\item The energy density is conserved at a local level; the total
energy density is therefore constant throughout the system.
\item The kinetic energy density is unique, it does not contain
any arbitrary terms.
\item The general exchange-correlation functional does not explicitly depend
on the density of charge.
\end{itemize}

From a practical point of view we note that self-consistency within
the LPS framework will be much faster to achieve, since the density
distribution is much more restricted. In particular the requirement
that the total energy density must be constant at every point of the
system should make the construction of fast algorithms quite easy.
In this respect it must also be noted that solutions of the equation
scale with the volume of the system, thus the number of atoms: the
method is therefore a true order-N method.

\section{Atomic systems}\label{atom_ana}

In the following we analyse the results of the LPS relation for
simple atomic systems containing one electron like H, He$^+$,
and Li$^{2+}$. Previously, we have found
that the direct consequence of the LPS equation is the energy
conservation on a local level.
\begin{equation}
\label{lps} \varepsilon_{tot}({\bf r})=\mu\rho({\bf r})=t({\bf r})+
v_{ext}({\bf r})\rho({\bf r})+v_H({\bf r})\rho({\bf r})+
v_{gxc}({\bf r})\rho({\bf r})
\end{equation}
As a next step, we analyse the potentials and energy terms which
contribute to the total energy,
\begin{equation}
\label{etot} E_{tot}=T+E_{ext}+E_H+E_{gxc}
\end{equation}
depending on an assumption about electron density. In general, the
electron distribution in atoms is spherically symmetric, with the
nucleus occupying the centre of the sphere. Unless otherwise stated
we use atomic units throughout the rest of the paper.

\subsection{Potentials and energy terms}

We assume in the following that the electron density decays
exponentially within the atom $\rho(r)=\alpha e^{-\beta r}$, with
$\alpha,\beta>0$. This type of electron density is exact for
hydrogen-like atoms, with the values (hydrogen) of
$\alpha=1/\pi,\beta=2$. With this ansatz all the relevant variables
of the atomic system can be explicitly written down. The number of
electrons is
\begin{equation}
N=\int d^3r\rho({\bf r})=4\pi\alpha\int_0^\infty dr r^2e^{-\beta r}=
\frac{8\pi\alpha}{\beta^3}
\end{equation}
The following quantities are necessary for evaluating the kinetic
energy density:
\begin{eqnarray}
\frac{[\nabla\rho(r)]^2}{\rho(r)}&=&\alpha\beta^2e^{-\beta r},
\nonumber \\
\nabla^2\rho(r)&=&\frac{d^2\rho(r)}{dr^2}+\frac{2}{r}\frac{d\rho(r)}{dr}
=\alpha e^{-\beta r}\left[\beta^2-\frac{2\beta}{r}\right]. \nonumber
\end{eqnarray}
Thus the kinetic energy density will be
\begin{eqnarray}
\label{t}
t(r)=-\frac{1}{4}\nabla^2\rho(r)+\frac{1}{8}\frac{[\nabla\rho(r)]^2}{\rho(r)}
=\alpha e^{-\beta
r}\left[-\frac{\beta^2}{8}+\frac{\beta}{2r}\right].\nonumber\\
\end{eqnarray}
The kinetic energy is
\begin{eqnarray}
\label{TT} T&=&\int d^3r t({\bf r})=4\pi\alpha\int_0^\infty dr
r^2e^{-\beta r}
\left[-\frac{\beta^2}{8}+\frac{\beta}{2r}\right]\nonumber\\&=&\pi\frac{\alpha}{\beta}
=\frac{N\beta^2}{8}.
\end{eqnarray}
The chemical potential is given by
\begin{equation}
\mu=\lim_{r\rightarrow\infty}\frac{t(r)}{\rho(r)}=
\lim_{r\rightarrow\infty}\left[-\frac{\beta^2}{8}+\frac{\beta}{2r}\right]
=-\frac{\beta^2}{8}
\end{equation}
The external potential is $v_{ext}(r)=-Z/r$, with $Z$ being the
nuclear charge; the electron-nuclear energy is consequently
\begin{eqnarray}
E_{ext}&=&\int d^3rv_{ext}({\bf r})\rho({\bf r})= -4\pi\alpha
Z\int_0^\infty dr re^{-\beta r}\nonumber\\&=&-\frac{4\pi\alpha
Z}{\beta^2} =-\frac{\beta NZ}{2}.
\end{eqnarray}
The Hartree potential can be obtained using Gauss' theorem via the
radial component of the electric field:
\begin{eqnarray}
E_r(r)&=&
\frac{4\pi\alpha}{r^2}\int_0^rdr'r'^2e^{-\beta r'}= \nonumber\\
&&-\frac{4\pi\alpha}{\beta^3r^2}[e^{-\beta r}(\beta^2r^2+2\beta r+2)-2],\nonumber\\
v_H(r)&=&-\int_\infty^r dr'E_{r'}(r')=\frac{4\pi\alpha}{\beta^3}
\left[\frac{2}{r}-e^{-\beta r}\left(\beta+\frac{2}{r}\right)\right]=
\nonumber \\ &&+\frac{N}{r}-e^{-\beta
r}\left[\frac{N\beta}{2}+\frac{N}{r}\right].\label{vh}
\end{eqnarray}
The Hartree energy is thus
\begin{eqnarray}
E_H&=&\int d^3rv_H({\bf r})\rho({\bf
r})\nonumber\\&=&4\pi\alpha\int_0^\infty dr r^2
\left[\frac{N}{r}-e^{-\beta
r}\left(\frac{N\beta}{2}+\frac{N}{r}\right)\right] \nonumber
\\&=&\frac{5\pi\alpha N}{2\beta^2}=\frac{5}{16}N^2\beta
\end{eqnarray}
The gxc-potential can be obtained from Eq.\ (\ref{lps}):
\begin{eqnarray}
v_{gxc}(r)&=&\mu-\frac{t(r)}{\rho(r)}-v_{ext}(r)-v_H(r)=-\frac{\beta^2}{8}+\frac{\beta^2}{8}\nonumber \\
&&-\frac{\beta}{2r}+\frac{Z}{r}
-\frac{N}{r}+e^{-\beta r}\left[\frac{N\beta}{2}+\frac{N}{r}\right] \nonumber\\
&=&\frac{1}{r}\left[Z-N-\frac{\beta}{2}\right]+ e^{-\beta
r}\left[\frac{N\beta}{2}+\frac{N}{r}\right]\label{vgxc}
\end{eqnarray}
Adding the Hartree potential to the gxc-potential gives us all
electron-electron related potentials,
\begin{equation}
\label{veff:4} v_{eff}(r)=v_H(r)+v_{gxc}(r)=\frac{2Z-\beta}{2r}.
\end{equation}
We get the gxc-energy as
\begin{eqnarray}
E_{gxc}&=&\int d^3rv_{gxc}({\bf r})\rho({\bf
r})=4\pi\alpha\int_0^\infty dr r^2 v_{gxc}(r)e^{-\beta r}=\nonumber
\\&&\frac{N\beta}{2}\left(Z-\frac{\beta}{2}-\frac{5}{8}N.\right)
\end{eqnarray}
According to Eq.\ (\ref{etot}) the total energy will then be
\begin{eqnarray}
E_{tot}&=& \frac{N\beta^2}{8}-\frac{\beta
NZ}{2}+\frac{5}{16}N^2\beta\nonumber\\
&&+\frac{N\beta}{2}\left(Z-\frac{\beta}{2}-\frac{5}{8}N\right)\nonumber
\\&=&-\frac{N\beta^2}{8}=\mu N.\label{etot1}
\end{eqnarray}
Comparing the total energy to the kinetic energy in Eq.\ (\ref{TT}),
we find that $E_{tot}=T+V=-T$. This demonstrates the validity of the
virial theorem for atomic systems, i.e.\ $2T+V=0$. Moreover, it
becomes clear that $\mu$ is an {\em average chemical potential},
i.e.\ $\mu=E_{tot}/N$, where $E_{tot}$ is the sum of the chemical
potentials as we build up our system from 1 to N electrons:
\begin{equation}
E_{tot}=\sum_{i=1}^N\mu_i\;\rightarrow\;\mu=\frac{\sum_{i=1}^N\mu_i}{N}.
\end{equation}
The single electron eigenvalues $\mu_i$ can directly be determined
by DFT. All terms in Eq.\ (\ref{lps}) are now parametrized. The
atomic systems are described by three parameters $Z,N,\beta$, although,
it has to be noted that the form of the electron
density considerably changes for $N>1$ which means the above procedure
is only valid for atomic systems with one electron.
$\beta$ is directly related to the total energy ($E_{tot}$), a fact
known from experiments and DFT calculations:
\begin{equation}
\beta=\sqrt{-\frac{8E_{tot}}{N}}.
\end{equation}
Now, we have all the tools to analyse the potentials for systems
containing one electron ($N=1$).

\subsection{H atom}

\begin{figure}[h]
\begin{center}
\includegraphics[width=\columnwidth]{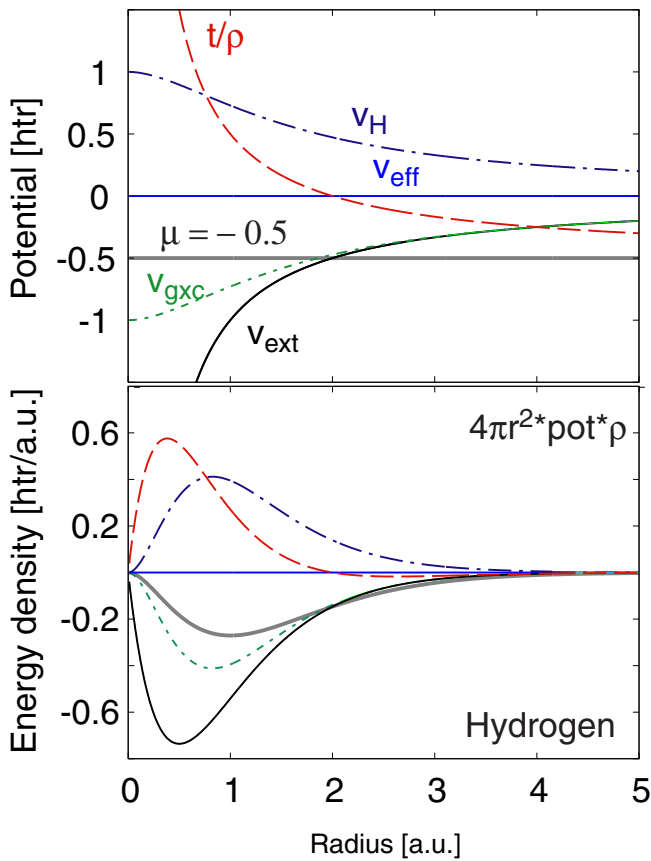}
\caption{Potentials (top) and energy densities (bottom) within the
hydrogen atom. Note that $v_{gxc}$ is equal, but of opposite sign as
$v_{H}$. The effective potential $v_{eff}$ is therefore zero.}
\label{fig:H}
\end{center}
\end{figure}

The electron distribution of hydrogen can of course be determined
analytically, using the single particle Schr\"odinger equation.
However, to elucidate the physical features of the LPS model, it is
instructive to choose a different route. In case of taking the
groundstate ($1s^1$) of the hydrogen the three parameters have the
following values: $Z=1,N=1,E_{tot}=-0.5$. The calculated potentials
and densities are shown in Fig.\ \ref{fig:H}.\\
\\
The calculated values $\alpha=1/\pi$ and $\beta=2$ are exactly the
same as in the analytical treatment. However, we gain additional
insights into the relation between Hartree and exchange-correlation
potentials: they are not zero, as they would be following the single
particle Schr\"odinger equation, but $v_{gxc}(r)=-v_H(r)$ at every
$r$. This behavior also defines the limit for the effective
potential $v_{eff}(r) = v_{gxc}(r)+v_H(r) \ge 0$. Physically
speaking, it indicates that a single electron does not interact with
itself, not, as in the conventional picture, because
electron-electron repulsion ($v_H$) is zero, but because it is
cancelled by exchange-correlation. For this reason we called
$v_{gxc}$ the {\em general exchange-correlation potential}, because
it is inconceivable, within a statistical many-electron treatment,
that it will be non-zero for a single electron. Another important
feature of $t/\rho$ is that it can be negative. However, this feature
was already analysed in detail by Sim et al. \cite{sim03}.

\subsection{Stability of atoms}

The mathematical condition for $v_{gxc}(r)=-v_H(r)$ can be
determined by comparing the terms in Eqs. (\ref{vh}) and
(\ref{vgxc}). It is $\beta=2Z$. Since, according to Eq.
(\ref{etot1}):
\begin{eqnarray}
E_{tot}=-\frac{N\beta^2}{8} = - \frac{N Z^2}{2} = \mu N \quad
\Rightarrow \quad \mu = -\frac{Z^2}{2},
\end{eqnarray}
which is always true for $N=1$. If the total energy $E_{tot}$
decreases to $E', E' < E_{tot}$, then the decay $\beta$, and,
consequently, the effective potential, will be:
\begin{eqnarray}
\beta &=& \sqrt{- 8 E'} > \sqrt{- 8 E_{tot}} \Rightarrow \\
&& v_{eff}({\bf r}) = v_{gxc}(r)+v_H(r) = \frac{1}{2r}\left(2 Z -
\beta\right) < 0.\nonumber
\end{eqnarray}
The energy $E_{tot}$ is therefore the minimum energy of the hydrogen
atom; the actual energy is bound from below. The atom is thus
stabilized by repulsive electron-electron interactions, which {\em
cannot} be more than compensated by exchange-correlation
interactions. A decrease of the energy, and thus an implosion of the
atom, is prevented by the fact that exchange-correlation would have
to be larger than the repulsive electron-electron interaction. The
physical cause, invoked here to explain atomic stability, is
different from the standard one, where it is thought that a decrease
of the atomic volume would increase the energy uncertainty of the
electron and thus increase the electron's energy. The standard
account is based on the Heisenberg uncertainty relations. The reason
we arrive at a different account within the LPS framework are the
local properties. Since within the standard
framework only global properties are accessible, an argument can
only be made on the basis of global properties, e.g. volume and
energy. However, the LPS framework provides direct access to all
{\em local} properties. Therefore the argument can be made on a
local basis, via the local interactions of electron density and
fields of interactions.

It is tempting to extend the analysis to the general case of
isolated electrons. In case of $N=1$ electron if $Z$ becomes very
small $Z \rightarrow 0$, the
effective potential in (\ref{veff:4}) will remain zero, because $2Z
- \beta = 0$. Electron charge therefore will become close to
homogeneous. But also in this case the effect of electron-electron
repulsion is cancelled because of exchange-correlation. This seems
to indicate, that regardless of the actual extension of electron
charge, electrons will {\em not} experience any repulsion between
the charge densities at different regions of the electron: electrons
are therefore stable entities.

In the general case $N \ge 1$ the total energy of an arbitrary
system of $N$ electrons  must be greater than
\begin{equation}
E_{tot} = - \frac{NZ^2}{2}.
\end{equation}
This value of $E_{tot}$ gives the exact total energy for a system of
non-interacting electrons. The total energy in this case cannot
decrease below $E_{tot}$ because this would correspond to an
attractive interaction between electrons.

Analysing $He^+$ and $Li^{2+}$ ions and the $2s^1$ excited state of
the electron in $H$ atom, all systems containing $N=1$ electron,
$v_{eff}=0$ was always found.

\subsection{Summary}

To sum up the results of this analysis of hydrogen-like systems:
(i) We found in all cases that the Hartree potential is non-zero.
This is in line with the general concept of DFT, where extended
distributions of electron charge always have an effect on
electrostatic potentials. (ii) We also found that the atoms are
stable due to the peculiar features of the general exchange
correlation potential: it is a negative contribution, equal in
magnitude to the Hartree potential, which prevents the repulsion
of electron charge to have any effect. The extent of the charge
distribution is thus bound from above. But it is also bound from
below, since the general exchange correlation potential does not
exceed the magnitude of the Hartree potential. Hydrogen-like atoms
as well as free electrons are therefore stable entities.

However, at this point the physical nature of this potential remains
mysterious: it does not explicitly depend on the charge
distribution, as has been generally derived from the LPS relation.
It can also not be related to exchange correlation in standard DFT,
since the systems are composed of only one electron.

\section{The origin of $v_{gxc}$}\label{gen_xc_pot}

At this point we have to look at the physical situation from a more
fundamental point of view. In general, electrons are described by
three physical properties: (i) Mass, (ii) charge, and (iii) spin.
From a fundamental point of view, one of these properties must be
related to the existence and the properties of the general exchange
correlation potential. It is straightforward to exclude charge
itself as its origin, since both effects of charge, the
electron-electron repulsion, and the electron-nuclear attraction,
are part of the description. Mass, even though it fits one of the
characteristics of the general exchange correlation, i.e. the
attractive nature, cannot be responsible for it, because the
coupling constant for gravitational interaction is tens of orders of
magnitude smaller than the coupling constant for electrostatic
interactions. This leaves only one viable option: the general
exchange correlation potential must be related to the magnetic
properties of electrons.

In general, motion of charge is related to the existence of
fields. Conversely, fields will affect the motion of charge. From
this perspective the motion of electron charge within a hydrogen
atom is likely to create corresponding fields, which shall have an
effect on the energy density at a particular point of the system.
There is, however, one restriction: these fields cannot be equal
to electromagnetic ${\bf E}$ and ${\bf B}$ fields, as this would
lead to a change of the energy of the atomic system due to
radiation. The field in question can therefore only be a vector
potential ${\bf A}$. The kinetic energy operator used in the
previous sections does not contain any field components. For this
reason it most likely gives only a limited account of the physical
environment.

\subsection{The modified LPS equation}

The existence of a field ${\bf A}$ and the ensuing kinetic energy
operator $\frac{1}{2}\left(-i\nabla - {\bf A}\right)^2$ will
introduce imaginary components into the Schr\"odinger equation. Real
functions $\phi = \rho^{1/2}$ are thus not sufficient to describe
electrons in this situation. This requires to generalize the
wavefunction for complex phases and to reinvestigate the LPS
relation according to this change. The motivation for this procedure
is to determine the physical origin of the general exchange
correlation potential. To this end we start with a free electron in
a three-dimensional potential well. In this case the electron can be
described as a plane wave,
\begin{equation}
\label{psi-free-electron} \Psi({\bf r})=\frac{1}{\sqrt{V}}e^{i{\bf
k}{\bf r}},
\end{equation}
with $V$ the volume of the potential well and ${\bf k}$ the momentum
of the electron. The above wavefunction results in a constant
electron density,
\begin{equation}
\rho({\bf r})=\Psi({\bf r})\Psi^*({\bf r})=\frac{1}{V}.
\end{equation}
From the above it seems to be clear that the LPS equation,
Eq.\ (\ref{lps_phi}) can not be complete, since for the free electron
the square root of the electron density is constant which implies
that $\mu$ has to be zero, which is obviously not the case.  A
similar remedy was suggested earlier by Wang and Carter, see section
VI.1 of \cite{wangcarter}. Here, we extend their results and assume
the most general wavefunction to be complex. The wavefunction
contains two important features of the electron: amplitude and
phase, generally
\begin{equation}
\Psi({\bf r},t)=\rho({\bf r},t)^{1/2}e^{i\varphi({\bf r},t)}.
\end{equation}
Writing the Schr\"odinger equation with this wavefunction in the
most general way (we omit space and time dependencies for brevity),
\begin{eqnarray}
i\frac{\partial\Psi}{\partial t}&=&\left[
-\frac{\partial\varphi}{\partial t}+ i\frac{1}{\rho^{1/2}}
\frac{\partial\rho^{1/2}}{\partial t}\right]\Psi=
\nonumber\\
\hat H\Psi&=&\left[-\frac{1}{2}\nabla^2+
v_{ext}+v_{H}+v_{gxc}\right]\Psi\,
\end{eqnarray}
results in two equations for the real and imaginary parts,
\begin{eqnarray}
-\frac{\partial\varphi}{\partial t}&=&-\frac{1}{2}
\frac{\nabla^2\rho^{1/2}}{\rho^{1/2}}+\frac{1}{2}(\nabla\varphi)^2
+v_{ext}+v_H+\Re e\; v_{gxc},\nonumber\\
\frac{1}{\rho^{1/2}}\frac{\partial\rho^{1/2}}{\partial t}&=&
-\frac{\nabla\rho^{1/2}\nabla\varphi}{\rho^{1/2}}-\frac{1}{2}\nabla^2\varphi
+\Im m\; v_{gxc}. \label{time_dep_ri}
\end{eqnarray}
Here, the most general $v_{gxc}$ is a complex potential. In a
stationary state an eigenvalue equation applies,
\begin{eqnarray}
\label{time_dep_gen} i\frac{\partial\Psi}{\partial t}=\left[
-\frac{\partial\varphi}{\partial t}+ i\frac{1}{\rho^{1/2}}
\frac{\partial\rho^{1/2}}{\partial t}\right]\Psi=\hat H\Psi=\mu\Psi,
\end{eqnarray}
which again results in two equations (real and imaginary parts),
\begin{eqnarray}
\label{lps-modified}
\mu&=&-\frac{1}{2}\frac{\nabla^2\rho^{1/2}}{\rho^{1/2}}+
\frac{1}{2}(\nabla\varphi)^2+v_{ext}+v_H+\Re e\; v_{gxc},\nonumber\\
0&=&-\frac{\nabla\rho^{1/2}\nabla\varphi}{\rho^{1/2}}-\frac{1}{2}\nabla^2\varphi
+\Im m\; v_{gxc}.
\end{eqnarray}
Eq.\ (\ref{time_dep_ri}) restricts the wavefunction to the simpler
form,
\begin{equation}
\label{psi-} \Psi({\bf r},t)=\rho({\bf r})^{1/2}e^{i\varphi({\bf
r})}e^{-i\mu t}= \psi({\bf r})e^{-i\mu t}.
\end{equation}
The first of Eqs.\ (\ref{lps-modified}) is the modified LPS equation
where the
phase function of the electron explicitly occurs modifying the
kinetic energy of the electron system (see next section). Applying
this equation to the free electron gives us the correct result for
the total energy ($\mu$ for $N=1$ electron):
\begin{equation}
\mu=\frac{1}{2}{\bf k}^2.
\end{equation}
This modification accounts for the well known problem in orbital
free methods, that the kinetic energy functionals of atomic systems,
usually described by some modification of the von Weizs\"acker
functional (see above), or the kinetic energy functionals of free
electron systems, like in many metals, are completely different.
Wang and Carter proposed a Lindhard function \cite{wangcarter},
which connects these two extreme cases in a common mathematical
model. By contrast, we find here that the difference can be
accounted for by the addition of a complex phase to the root of the
charge density.

We can sum up the results of this analysis by saying that it is
essential, for the description of free electrons, to take into
account both the amplitude and the phase information. As a
consequence of this, the LPS equation has to be modified. The phase
information, we found, enters as an additional term in the kinetic
energy density, $(\nabla\varphi)^2/2$.

\subsection{The physical meaning of general exchange correlations}

In previous sections we have shown that it is essential, for the
stability of atoms, that the general exchange correlation is not
zero. We claim that this is also the case for the free electron, see
next section. Furthermore, it was shown that a complex phase is
essential for obtaining the correct energy eigenvalue. Here, we want
to determine the origin of this phase and show that the general
exchange correlation is due to an inertial field of the electron. To
this end let us consider the modified momentum of a charged particle
travelling through a region of space with nonzero electromagnetic
potentials. It is well known that the momentum of such an electron
is
\begin{equation}
\hat p=-i\nabla-{\bf A},
\end{equation}
where ${\bf A}$ is the vectorpotential, while the curl of the
vectorpotential is the external magnetic field,
\begin{equation}
{\bf B}=\nabla\times{\bf A}.
\end{equation}
The Aharonov-Bohm effect \cite{aharonov} established the importance
of electromagnetic potentials in quantum physics. For example, in a
system containing an external magnetic field, e.g. a solenoid, an
electron is affected by the vectorpotential, even propagating
through a region of space, where the external magnetic field is
zero. It acquires the phase
\begin{equation}
\varphi_{AB}=\int\limits_{path}{\bf A}({\bf r})d{\bf r}.
\end{equation}
The effect is also observed in experiments
\cite{webb,schwarzschild}. Under the condition that an inertial
field ${\bf A}({\bf r})$, i.e. a field, which is {\em not} due to
external sources, but due to the {\em propagation of the electron
itself}, exists, the properties of the general exchange-correlation
potential can be readily derived. The Hamiltonian now takes the
form,
\begin{eqnarray}
\hat H&=&\frac{1}{2}[-i\nabla-{\bf A}({\bf r})]^2+v_{ext}({\bf
r})+v_H({\bf r})\\&=&-\frac{1}{2}\nabla^2+\frac{1}{2}{\bf A}({\bf
r})^2+\frac{i}{2}[{\bf A}({\bf r})\nabla+\nabla{\bf A}({\bf
r})]\nonumber\\&&+ v_{ext}({\bf r})+v_H({\bf r}).\nonumber
\end{eqnarray}
Comparing this to Eq.\ (\ref{lps_phi}) and using that
$v_{eff}=v_H+v_{gxc}$ it is straightforward to
conclude that the generalized exchange-correlation operator is
\begin{eqnarray}
\label{v_gxc-A} \hat v_{gxc}({\bf r})&=&\frac{1}{2}{\bf A}({\bf
r})^2+ \frac{i}{2}[{\bf A}({\bf r})\nabla+\nabla{\bf A}({\bf
r})]\nonumber\\&=& \frac{1}{2}{\bf A}({\bf r})^2+
\frac{i}{2}[\nabla\cdot{\bf A}({\bf r})+2{\bf A}({\bf r})\nabla].
\end{eqnarray}
Using this notation the Hamiltonian can be written as
\begin{equation}
\hat H=-\frac{1}{2}\nabla^2+\hat v_{ext}({\bf r})+\hat v_H({\bf r})+
\hat v_{gxc}({\bf r}).
\end{equation}
The actions of each term of the Hamiltonian on the time-space
separable wavefunction, Eq.\ (\ref{psi-}) are
\begin{eqnarray}
-\frac{1}{2}\nabla^2\Psi({\bf r},t)&=&\left(
-\frac{1}{2}\frac{\nabla^2\rho({\bf r})^{1/2}}{\rho({\bf r})^{1/2}}+
\frac{1}{2}[\nabla\varphi({\bf r})]^2\right)\Psi({\bf
r},t)\nonumber\\&&-i\left( \frac{1}{2}\nabla^2\varphi({\bf r})+
\frac{\nabla\rho({\bf r})^{1/2}}{\rho({\bf
r})^{1/2}}\nabla\varphi({\bf r})
\right)\Psi({\bf r},t),\nonumber\\
\hat v_{ext}({\bf r})\Psi({\bf r},t)&=&v_{ext}({\bf r})\Psi({\bf r},t),\\
\hat v_H({\bf r})\Psi({\bf r},t)&=&v_H({\bf r})\Psi({\bf r},t),\nonumber\\
\hat v_{gxc}({\bf r})\Psi({\bf r},t)&=&\left(\frac{1}{2}{\bf A}({\bf
r})^2 -{\bf A}({\bf r})\nabla\varphi({\bf r})\right)\Psi({\bf
r},t)\nonumber\\&&+ i\left(\frac{\nabla\cdot{\bf A}({\bf r})}{2}+
\frac{\nabla\rho({\bf r})^{1/2}}{\rho({\bf r})^{1/2}}{\bf A}({\bf
r}) \right)\Psi({\bf r},t).\nonumber
\end{eqnarray}
It is important to mention that the kinetic energy is modified by a
term, $[\nabla\varphi({\bf r})]^2/2$ which directly relates to the
phase of the electron. This differs from the conception of e.g.
Vignale and Kohn \cite{vig96}, where the field is related to current
density. In previous work on orbital free DFT this term was never
taken into account in the way described here. Adding all terms
yields two equations for real and imaginary part, respectively,
\begin{eqnarray}
\label{lps-A} \mu&=&-\frac{1}{2}\frac{\nabla^2\rho({\bf
r})^{1/2}}{\rho({\bf r})^{1/2}}+
\frac{1}{2}[{\bf A}({\bf r})-\nabla\varphi({\bf r})]^2+v_{ext}({\bf r})+v_H({\bf r}),\nonumber\\
0&=&\left(\frac{1}{2}\nabla+\frac{\nabla\rho({\bf
r})^{1/2}}{\rho({\bf r})^{1/2}}\right) [{\bf A}({\bf
r})-\nabla\varphi({\bf r})].
\end{eqnarray}
The second equation leads to the simple solution
\begin{equation}
\label{a-gradphi} {\bf A}({\bf r})=\nabla\varphi({\bf
r})\quad\rightarrow\quad \varphi({\bf r})=\int\limits_{{\bf
r}_0}^{{\bf r}}{\bf A}({\bf s})d{\bf s},
\end{equation}
where $\varphi({\bf r}_0)=0$. The phase of the electron is therefore
related to a quantity which can be interpreted as a vectorpotential.
The result seems not surprising, considering the Aharonov-Bohm
effect \cite{aharonov}. However, an objection may be raised that this
effect relates external ${\bf A}$ fields to observed phase shifts,
and that a phase-shift due to propagation of the electron itself will
change the theoretical predictions for the experiments. The readers
may convince themselves that this is not the case. If two electrons
passing on either side of a solenoid possess the same inertial field
${\bf A}$, then the phase difference between any two paths with the
same endpoint around the solenoid will be given by
\begin{equation}
\Delta\varphi_{AB}=\oint{\bf A}_{ext}({\bf r})d{\bf r}+
\oint{\bf A}({\bf r})d{\bf r}=\int{\bf B}_{ext}({\bf r})d{\bf F}+0,
\end{equation}
it is therefore determined exclusively by the external field. The
main difference in our present model is that ${\bf A}({\bf r})$ is
the inertial field due to electron propagation, not a
vectorpotential caused by external fields. The magnetic field due to
this vectorpotential will vanish, since the curl of a gradient is
always zero. Substituting Eq.\ (\ref{a-gradphi}) into the first of
Eqs.\ (\ref{lps-A}) gives
\begin{equation}
\label{lps-original} \mu=-\frac{1}{2}\frac{\nabla^2\rho({\bf
r})^{1/2}}{\rho({\bf r})^{1/2}}+ v_{ext}({\bf r})+v_H({\bf r}).
\end{equation}
The electron wavefunction takes the form,
\begin{equation}
\Psi({\bf r},t)=\rho({\bf r})^{1/2} e^{i\int_{{\bf r}_0}^{{\bf
r}}{\bf A}({\bf s})d{\bf s}}e^{-i\mu t}.
\end{equation}
Using this solution, we find that the real kinetic energy density
$t_R$ is not the bosonic kinetic energy density, employed in the
original LPS relation, but that it possesses an additional term,
depending on the inertial field ${\bf A}$.
\begin{eqnarray}
\Re e\; t_R({\bf r})&=&-\frac{1}{2}\rho({\bf
r})^{1/2}\nabla^2\rho({\bf r})^{1/2}+ \frac{1}{2}\rho({\bf r}){\bf
A}({\bf r})^2, \\ \Im m\; t_R({\bf r})&=&\frac{1}{2}\rho({\bf
r})\nabla{\bf A}({\bf r})+ \rho({\bf r})^{1/2}\nabla\rho({\bf
r})^{1/2}{\bf A}({\bf r}).\nonumber
\end{eqnarray}
In this case we obtain a corresponding exchange correlation energy
density described by:
\begin{eqnarray}
\Re e\;\varepsilon_{gxc}({\bf r})&=&-\frac{1}{2}\rho({\bf r}){\bf
A}({\bf r})^2, \nonumber \\ \Im m\;\varepsilon_{gxc}({\bf r})&=&-\Im
m\;t_R({\bf r}).
\end{eqnarray}
If the inertial field $A({\bf r})$ vanishes, the exchange
correlation potential $\;\varepsilon_{gxc}({\bf r})$ is zero. In
addition, the sum of the real kinetic energy density $t_R$ and the
exchange correlation energy density is always equal to the bosonic
kinetic energy density:
\begin{equation}
\ \Re e\;t({\bf r})+\Re e\;\varepsilon_{gxc}({\bf r})=
-\frac{1}{2}\rho({\bf r})^{1/2}\nabla^2\rho({\bf r})^{1/2}.
\end{equation}
This entails that the fermi-ionic nature of electrons is due to
their inertial fields. We shall show further down that a gauge
transformation, which eliminates the inertial fields, reverts the
problem back to a problem of interacting bosons.

\subsection{Free electrons}

To demonstrate the consequences of the framework developed, let us
analyse a free electron enclosed in a finite volume $V$. The problem
is quite interesting to analyse, as DFT does not describe the system
correctly. This is due to the fact that in DFT both exchange
correlation potential $v_{xc}$ and Hartree potential $v_H$ vanish,
which is at odds with the physical situation comprising a finite
distribution of electron charge. In our description we treat the
Hartree term in the correct manner. From general considerations we
have found that $v_{gxc}=-v_H$ for systems of one electron. It
should also be noted that $v_{ext}$ is zero within the box and
infinity outside. As the phase of the wavefunction in Eq.\
(\ref{psi-free-electron}) is $\varphi({\bf r})={\bf k}{\bf r}$, the
inertial field of the free electron is
\begin{equation}
{\bf A}({\bf r})=\nabla\varphi({\bf r})={\bf k}.
\end{equation}
This means that the inertial field is just the momentum of the free
electron system which is uniform within the volume. This feature of
free electrons was predicted by heuristic arguments some years ago
\cite{wah1998}. The curl of the inertial field is, of course, zero.
The magnetic field therefore vanishes. Taking this fact and the
explicit form of $v_{gxc}$ in Eq.\ (\ref{v_gxc-A}) into account, we
can write the modified LPS equation as
\begin{eqnarray}
&&\left[-\frac{1}{2}\nabla^2+\hat v_H({\bf r})+ \hat v_{gxc}({\bf
r})\right]\frac{1}{\sqrt{V}}e^{i{\bf k}{\bf r}}\\&=&
\left[\frac{1}{2}{\bf k}^2+v_H({\bf r})-\frac{1}{2}{\bf k}^2
\right]\frac{1}{\sqrt{V}}e^{i{\bf k}{\bf r}}=
\mu\frac{1}{\sqrt{V}}e^{i{\bf k}{\bf r}}.\nonumber
\end{eqnarray}
Since $v_{eff} = v_H - {\bf k}^2/2 = 0$, the chemical
potential $\mu={\bf k}^2/2$, and all potentials and also the kinetic
energy density divided by the density have the same magnitude:
\begin{equation}
\frac{t}{\rho}=v_H=-v_{gxc}=\frac{1}{2}{\bf k}^2=\mu.
\end{equation}
In addition, they are also equal to the components of the total
energy of the system, as $N=1$:
\begin{equation}
\mu=E_{tot}=T=E_H=-E_{gxc}.
\end{equation}
As $E_H+E_{gxc}=0$, $E_{tot}=T$ and $V=0$. However, this is not the
general formulation of the problem. In the general case, elaborated
in detail for a homogeneous electron gas in the next section, the
field ${\bf A}({\bf r}) = {\bf A}_r({\bf r}) + i {\bf A}_i({\bf r})$
will be complex, and we have to account for partial waves ${\bf
A}^+$ and ${\bf A}^-$ traveling in opposing directions. The general
treatment will be based on a modification of the LPS equation to
account for this situation.

\subsection{Electrons in hydrogen}

Let us determine the inertial field of the electron in a
hydrogen-like atom. The groundstate density and wavefunction can be
written as
\begin{equation}
\rho(r)=\frac{Z^3}{\pi}e^{-2Zr}\quad;\quad\psi(r)=
\left(\frac{Z^3}{\pi}\right)^{1/2}e^{-Zr}.
\end{equation}
The system has a total energy of $-Z^2/2$. Acting $\hat v_{gxc}$ in
Eq.\ (\ref{v_gxc-A}) on this wavefunction results in two equations
for the real part and imaginary part, respectively,
\begin{eqnarray}
\Re e\; v_{gxc}&=&\frac{1}{2}A_r(r)^2,\nonumber \\ \Im m\;
v_{gxc}&=&\frac{\nabla\cdot{\bf A}(r)}{2}-ZA_r(r)=0.
\end{eqnarray}
Here, we assumed that ${\bf A}(r)$ has only radial components,
$A_r(r)$. We have already calculated $v_{gxc}(r)$ in Eq.\
(\ref{vgxc}), the parameters in our case are $N=1$ and $\beta=2Z$.
Comparing this to the real part of $v_{gxc}$, the inertial field can
be obtained,
\begin{eqnarray}
\frac{1}{2}A_r(r)^2=-\frac{1}{r}+e^{-2Zr}\left[Z+\frac{1}{r}\right],
\nonumber\\
A_r(r)=\pm\sqrt{-\frac{2}{r}+2e^{-2Zr}\left[Z+\frac{1}{r}\right]}.
\end{eqnarray}
The inertial field $A$ in this case is imaginary. However, this
results is only puzzling, as long as the direction of motion remains
unconsidered. We have found, for the free electron, that $A$ is a
vector field, which is parallel to the motion vector of the
electron. If we assume, in case of the hydrogen electron, that the
vector of motion is radial, then it cannot be unique, because in
this case the electron distribution cannot be stable. This leads to
the conclusion, that the standard time-independent solution must
comprise two separate cases, with the vector of motion either
parallel, or anti-parallel to the radial vector. The field $A$
should therefore be a superposition of two partial fields, $A^+$,
and $A^-$. With the ansatz for the partial fields,
\begin{equation}
A_r^+(r) = e^{i \chi(r)} \qquad A_r^-(r) = - e^{-i \chi(r)},
\end{equation}
we obtain for the total field $A_r(r)$
\begin{equation}
A_r(r) = A_r^+(r) + A_r^-(r) = 2 i \sin \chi(r).
\end{equation}
The inertial field in hydrogen is thus described by a radial
function $\chi(r)$, which complies with
\begin{equation}
\chi(r) = \arcsin
\left(\pm\frac{1}{2}\sqrt{\frac{2}{r}-2e^{-2Zr}\left[Z+\frac{1}{r}\right]}\right).
\end{equation}

\begin{figure}[h]
\begin{center}
\includegraphics[width=\columnwidth]{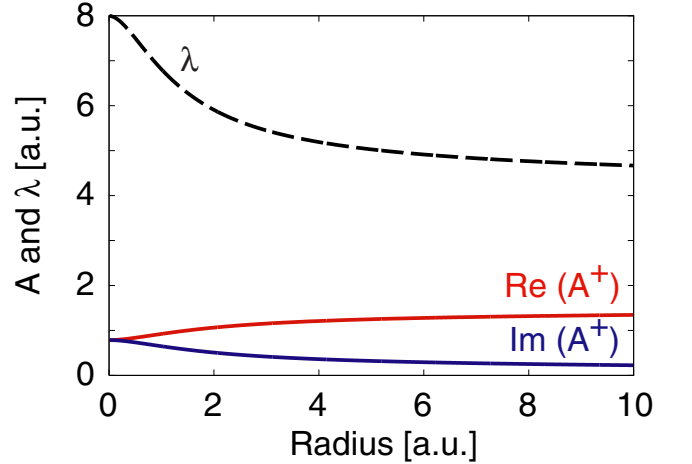}
\caption{Real (red graph) and imaginary (blue graph) components of
the inertial field $A_r^+(r)$ within hydrogen. The components are
equal at the limit of zero radius. The wavelength of the hydrogen
electron (dashed graph) saturates at large radii.} \label{fig:IF}
\end{center}
\end{figure}

For $Z \le 2$ this leads to a well behaved solution, since
\begin{equation}
e^{-2 Z r} = \sum_{n=0}^{\infty}\frac{(-2Zr)^n}{n!} = 1 - 2Zr +
2Z^2r^2 - ...
\end{equation}
The limit for $r \rightarrow 0$ is consequently:
\begin{equation}
\lim_{r \rightarrow
0}\sqrt{\frac{2}{r}-2e^{-2Zr}\left[Z+\frac{1}{r}\right]} =
\sqrt{2Z}.
\end{equation}
Note that the real components of $A_r^+(r)$ and $A_r^-(r)$ possess
opposite signs, in line with the previous findings for free
electrons. The real and imaginary components of $A_r^+(r)$ are
shown in Fig. \ref{fig:IF}. It is interesting to note that
$A_r^+(r)$ (i) does not show a singularity like the external
potential, and (ii) does not decay exponentially. If we assume
that the real part of the inertial field is related to electron
motion, then the differential of the phase $\nabla \phi = \Re e\;
A_r^+(r)$ can be used to calculate the phase. Since the phase is
related to the wavelength of the electron via $dr/\lambda =
d\phi/2 \pi$, we can also calculate the wavelength of the hydrogen
electron as a function of radius:
\begin{equation}
\lambda(r) = \frac{dr}{d\phi(r)} 2 \pi.
\end{equation}
As seen in Fig. \ref{fig:IF} the wavelength near the hydrogen core
is about eight or $5\pi/2$ atomic units, it saturates for high
radii at about $3\pi/2$.
From a physical point of view, the real part of $A$ increases as the
wavelength decreases. By contrast the imaginary part of $A$
decreases with decreasing density and increasing radius.
Tentatively, one could thus interpret the real part of $A$ as a
physical quantity related to motion, and the imaginary part of $A$
as something akin to internal friction of the electron. At the
boundary of the nucleus, or for $r \rightarrow 0$, both components
are equal, signifying that at this point the energy of motion is
completely transformed into friction.

\subsection{A system of $N$ free electrons: the Pauli principle}

Here, we demonstrate that the inertial field also accounts for the
Pauli exclusion principle in a natural manner. Separate Pauli
potentials, as proposed by March, or Levy and Ou-Yang
\cite{march85,levy88} are therefore unnecessary. To this end we
consider the simplest possible system, the system of free electrons.
The inertial field of a single electron is given by ${\bf A}_1$, the
corresponding energy of the electron is
\begin{equation}
E_1 = \frac{1}{2}{\bf A}_1^2 = \frac{1}{2}{\bf k}_1^2.
\end{equation}
In a system of two electrons the two inertial fields will be
superimposed. Assuming, initially, that superimposed fields are
parallel in space, the combined field of the two electrons will be
\begin{equation}
{\bf A}_2 = 2 {\bf A}_1 \quad \Rightarrow \quad E_2 = \frac{1}{2}
2^2 {\bf A}_1^2 = 4 E_1.
\end{equation}
Since the density of charge will increase by a factor of two, the
Hartree potential and the general exchange correlation potential
will increase by the same amount: a system of two electrons remains
thus stable due to the superposition of the two inertial fields. A
different way of expressing this situation would be that for every
single electron the effective potential remains zero. For a system
of $N$ electrons an equivalent description holds, so that the total
energy of $N$ electrons is given by:
\begin{equation}
E_N = N^2 E_1.
\end{equation}
The result is exactly the same as the one derived for a free
electron gas in one dimension, which is usually accomplished using
periodic boundary conditions. Here, we have only used the property
of any field, i.e. the superposition of amplitudes.

However, in a general three dimensional system the direction of
fields can vary. In this case the assumption of parallel vector
potentials is not the state of minimum energy. Then the state of
minimum energy is described by:
\begin{equation}
{\bf A}_N = \sum_{i=1}^N {\bf A}_i \qquad {\bf A}_N^2 = \mbox{min}.
\end{equation}
Since the individual fields ${\bf A}_i$ will have equal length
$|{\bf A}_i| = |{\bf k}_1|$, the minimum of the total field ${\bf
A}_N$ will be a sphere, equivalent to the Fermi sphere in a three
dimensional free electron model. Note that in all these cases the
addition of one electron increases the volume of occupied states in
reciprocal space according to the Pauli exclusion principle. The
volume of one cell in this sphere is determined by the amplitude of
the vector potential.

\subsection{Gauge transformations}

In this section we assume a stationary electron state and
investigate how it is transformed applying a gauge transformation.
The effect on the Schr\"odinger equation is also determined. This
analysis shall aid us in developing methods for an effective
solution of the electron problem. Here, the aim is to find a
suitable gauge for the inertial field ${\bf A}$. Electromagnetic
potentials can be transformed with a gauge transformation which
results in the same electromagnetic fields,
\begin{eqnarray}
{\bf A}'({\bf r})&=&{\bf A}({\bf r})-\nabla\Gamma({\bf r},t),\nonumber\\
v_{el.scal}({\bf r})'&=&v_{el.scal}({\bf r})+
\frac{\partial\Gamma({\bf r},t)}{\partial t},\\
{\bf E}'={\bf E}&;&{\bf B}'={\bf B}.
\end{eqnarray}
with $v_{el.scal}=v_{ext}+v_H$ the electromagnetic scalar potential
and $\Gamma$ the gauge function. The wavefunction transforms like
\begin{equation}
\Psi'({\bf r},t)=\Psi({\bf r},t)e^{-i\Gamma({\bf r},t)}.
\end{equation}
It is easy to show that the Schr\"odinger equation transforms
according to the following rule,
\begin{eqnarray}
\hat H\Psi({\bf r},t)&=&\mu\Psi({\bf r},t)\;\rightarrow\\
\hat H'\Psi'({\bf r},t)&=&i\frac{\partial\Psi'({\bf r},t)}{\partial
t}= \left[\mu+\frac{\partial\Gamma({\bf r},t)}{\partial t}\right]
\Psi'({\bf r},t),\nonumber
\end{eqnarray}
where the initial state $\Psi({\bf r},t)=\psi({\bf r})e^{-i\mu t}$
is supposed to be stationary. Gauge transformations are potentially
useful in reducing the complexity of the problem and finding general
solutions. For example, the following transformation reverts a
general wavefunction back to the root of the charge density,
\begin{eqnarray}
\Gamma({\bf r},t) &=& \int_{{\bf r}_0}^{{\bf r}} {\bf A}({\bf s})
d{\bf s} - \mu t \quad \Rightarrow \quad {\bf A}' = {\bf A} - \nabla
\Gamma = 0,\nonumber\\
\Psi'({\bf r}) &=& \rho^{1/2}({\bf r}),
\end{eqnarray}
which describes a bosonic system as
\begin{equation}
\left[-\frac{1}{2}\nabla^2+v_{ext}+v_H-\mu\right]\rho^{1/2}=
i\frac{\partial\rho^{1/2}}{\partial t}=0.
\end{equation}
This is the original LPS equation, see Eq.\ (\ref{lps_phi}), without
exchange correlation potentials. It describes interacting bosons
($v_{eff}=v_H
> 0$), and does not comply with the Pauli principle. The obvious
conclusion from this result is that the Pauli principle and the
fermionic character of electrons is due to their phases.

\section{The general problem}\label{gen_prob}

In previous sections it was established that the general problem
of finding the groundstate of an $N$-electron system can be
described as finding the six components - three real, three
imaginary - of the complex vector field ${\bf A}({\bf r})$, the
distribution of electron density $\rho({\bf r})$, and the chemical
potential $\mu$. The complex vector field determines the phase of
electrons, which is {\em not} described by charge density alone.
From this perspective the seeming detour taken in Kohn-Sham DFT is
quite understandable. Since charge density alone is insufficient
to guarantee phase-continuity throughout a system, it has to be
modelled by continuous phases of single electron states, taking
into account electron-electron interactions via
exchange-correlation functionals.

Within the present context, this detour is unnecessary, if the
inertial field can be calculated. That in this case six additional
values have to be determined for every point of the system is not
a critical problem. After all, the whole computation still scales
linearly, even though the prefactor might be somewhat larger. In
addition, boundary conditions apply for the inertial field, and
the chemical potential has to be constant throughout the system.
Given a complex vector field ${\bf A}({\bf r}) = {\bf A}_r({\bf
r}) + i {\bf A}_i({\bf r})$, the general problem can be formulated
in two equations:
\begin{eqnarray}
\frac{1}{2}\left[\left(-\nabla^2 + {\bf A}_r^2 - {\bf A}_i^2 +
{\bf A}_i\nabla - \nabla {\bf A}_i\right) \right.\nonumber
\\
\left.+ v_{ext} + v_H -
\mu\right]\rho^{1/2} = 0,\nonumber \\
\left(\nabla{\bf A}_r - {\bf A}_r\nabla - 2{\bf A}_r{\bf
A}_i\right)\rho^{1/2} = 0.
\end{eqnarray}
It is straightforward to separate the different contributions and
to assign them to potentials within standard DFT. The square of
the real part of the vector field accounts for the Pauli
principle, as discussed earlier. Thus we may define a Pauli
potential by:
\begin{equation}
v_P({\bf r}) = \frac{1}{2}\left[{\bf A}_r^2({\bf r}) \right].
\end{equation}
The exchange and correlation potentials are defined in standard
DFT as the difference of the kinetic energy between interacting
and non-interacting electrons. They correspond in this picture to
an operator, $\hat{v}_{XC}({\bf r})$, described by:
\begin{equation}
\hat{v}_{XC}({\bf r}) = \frac{1}{2}\left[- {\bf A}_i^2({\bf r}) +
{\bf A}_i({\bf r})\nabla - \nabla {\bf A}_i({\bf r})\right].
\end{equation}
The Pauli potential and the exchange-correlation potential are not
independent from each other, since they have to comply with the
auxiliary condition:
\begin{equation}\label{aux:eq}
\left[\nabla{\bf A}_r({\bf r}) - {\bf A}_r({\bf r})\nabla - 2{\bf
A}_r({\bf r}){\bf A}_i({\bf r})\right]\rho^{1/2}({\bf r}) = 0.
\end{equation}
Given this condition, the general problem is thus described in a
similar fashion to the original LPS equation by (the dependency on
coordinates is again omitted for brevity):
\begin{equation}\label{gen:eq}
\left[-\frac{1}{2}\nabla^2 + \hat{v}_{XC} + v_P + v_{ext} + v_H -
\mu\right]\rho^{1/2} = 0.
\end{equation}
The full potential therefore has four distinct terms: (i) an
external potential $v_{ext}$, which only depends on the
distribution of positive charge, it is described in every DFT
framework; (ii) a Hartree potential $v_H$, which depends only on
the charge distribution, which also is present in standard DFT
methods; (iii) a Pauli potential $v_P$, which depends on the
amplitude of the inertial field, in standard Kohn-Sham theory it
is accounted for by computing the solutions for single electron
states; and (iv) an exchange-correlation potential $\hat{v}_{XC}$,
which depends on the phase of the inertial field, described by
local densities and their gradients in standard methods. Note that
within the present framework many-body effects are related to the
inertial field ${\bf A}$; in general they are thus mediated by
field interactions. The effective potential, used in the original
LPS relation, is thus in fact an operator, given by:
\begin{equation}
\hat{v}_{eff}({\bf r}) = \hat{v}_{XC}({\bf r}) + v_P({\bf r}) +
v_H({\bf r}).
\end{equation}

\subsection{Homogeneous electron gas}

As an example, let us discuss the solution for a homogeneous
electron gas. From the auxiliary equation Eq. \ref{aux:eq}, we may
infer, as the simplest solution, that the real part of the vector
potential is zero:
\begin{equation}
{\bf A}_r({\bf r}) = {\bf A}_r^+({\bf r}) + {\bf A}_r^-({\bf r}) =
0.
\end{equation}
From translational symmetry for the external potential and the
Hartree potential, and from the fact that the Hartree potential
scales with the density of charge, as well as from the
translational invariance of the charge density $\rho({\bf r}) =
\rho_0$, we get:
\begin{eqnarray}
v_H({\bf r}) &=& \int d^3{\bf r}' \frac{\rho_0}{|{\bf r}-{\bf r}'|}
= \rho_0 \int d^3{\bf r}' \frac{1}{|{\bf r}-{\bf r}'|} = \rho_0
\alpha\nonumber\\
v_{ext} + v_H &=& - V_0 + \alpha \rho_0.
\end{eqnarray}
Thus we arrive at the nonlinear equation for the imaginary part of
the potential, or:
\begin{equation}
\frac{1}{2}\left[- {\bf A}_i^2({\bf r}) - \nabla {\bf A}_i({\bf
r})\right] + \alpha \rho_0 - V_0 = \mu.
\end{equation}
Choosing again, for translational invariance, a solution where
${\bf A}_i^2({\bf r}) = A_i^2$, we finally arrive at:
\begin{equation}
\mu = - \frac{A_i^2}{2} + \alpha \rho_0 - V_0.
\end{equation}
For a neutral system, where the external potential due to the
positive background charge and the Hartree potential must be
equal, or $\alpha \rho_0 - V_0 = 0$, this amounts to:
\begin{equation}
\mu = -\frac{A_i^2}{2}.
\end{equation}
In this case the exchange-correlation potential of the homogeneous
electron gas is described by $-A_i^2/2$. From a known chemical
potential $\mu$ the inertial potential can therefore be calculated
in a straightforward manner. Under the same conditions the vector
field for slowly varying densities, which we denote by ${\bf
A}_h$, will be given by (we denote by $\mu_h$ the chemical
potential of the homogeneous electron gas of a specific density
$\rho$):
\begin{equation}\label{ima:eq}
i {\bf A}_h({\bf r}) = \pm i \sqrt{-2 \mu_h[\rho({\bf r})]}\,
\frac{\nabla \mu_h[\rho({\bf r})]}{\left|\nabla \mu_h[\rho({\bf
r})]\right|}.
\end{equation}
Consistent with earlier findings, the exchange-correlation
potential does not depend explicitly on the density of charge. The
explicit Pauli potential for the $N$ electrons of the homogeneous
electron gas can be set to zero, because it is implicitly
contained in the chemical potential $\mu_h$. $\mu_h[\rho]$, in
turn, can be determined by standard methods in DFT.

\subsection{Local density approximation}

If the vector field is calculated from the homogeneous electron
gas, then the missing terms in the general problem are accounted
for. This amounts to a local density approximation for the general
problem. Assuming a distribution of $M$ ions, and a number of $N$
electrons within our system, and with ${\bf A}_h({\bf r})$ from
Eq. \ref{ima:eq}, the general problem in the local density
approximation is described by:
\begin{eqnarray}
\mu &=& -\frac{1}{4}\frac{\nabla^2\rho({\bf r})}{\rho({\bf r})}+
\frac{1}{8}\frac{[\nabla\rho({\bf r})]^2}{\rho({\bf r})^2} -
\frac{1}{2}{\bf A}_h^2({\bf r})
\nonumber \\
&&\pm \frac{1}{2 \rho^{1/2}({\bf r})}\left[{\bf A}_h({\bf
r})\nabla - \nabla {\bf A}_h({\bf r})\right] \rho^{1/2}({\bf r})
\nonumber \\
&& - \sum_{i=1}^M\frac{Z_i}{|{\bf r}-{\bf R}_i|} +\int
d^3r'\frac{\rho({\bf r}')}{|{\bf r}-{\bf r}'|}, \\
{\bf A}_h({\bf r}) &=& \sqrt{-2 \mu_h[\rho({\bf r})]}\, \frac{\nabla
\mu_h[\rho({\bf r})]}{\left|\nabla \mu_h[\rho({\bf r})]\right|},
\nonumber \\ N &=& \int d^3r' \rho({\bf r}'). \nonumber
\end{eqnarray}
It is interesting to note that the exchange-correlation potential
contains not only the values of ${\bf A}_h({\bf r})$ and
$\rho({\bf r})$, but also their derivatives. This is due to the
fact that the inertial fields are related to the phases of the
electrons.

In this case all terms of the equation depend on the local charge
and its derivatives. A self-consistency cycle then can proceed in
the familiar manner by iterating the charge density distribution
until the chemical potential is a minimum. As the only input
parameter in the calculation is the density of charge itself, with
the auxiliary quantities taken from the results obtained for a
homogeneous electron gas, the method is a true order-$N$ method.

However, it is not strictly necessary to remain at this level of
approximation. Assume, that a groundstate solution has been found.
Then the initial values of $\rho({\bf r})$, ${\bf A}_i({\bf r})$,
and $\mu$ are known. Subsequently, the solution can be refined, by
(i) using Eq. \ref{aux:eq} to determine the real part of the
vector potential; (ii) solving the general problem with the help
of Eq. \ref{gen:eq}; and (iii) iterating the vector field and the
density distribution until the chemical potential is a minimum. In
this case the solution is the true many-body solution of the
problem.

\section{Summary and discussion}\label{sum_dis}

Let us briefly summarize the findings in the previous sections. It
was found that the LPS relation describes local conservation of
energy density. It can be generally derived by variational methods
from the total energy functional, and emerges as the
Euler-Lagrange equation of the energetic minimum. Its application
to hydrogen-like atoms showed that these systems are stable,
because the effective potential, $v_{eff}=v_H + v_{gxc}$, is zero.
However, the general exchange correlation potential, $v_{gxc}$,
which is equal in magnitude but of opposite sign as the Hartree
potential, cannot be explained from standard exchange and
correlation effects for systems containing only one electron. An
analysis with a more general kinetic operator including the effect
of ${\bf A}$ fields showed that the gxc-potential is due to an
inertial field ${\bf A}$, related to electron motion, which leads
to a lowering of total energy. The phase of free electron is also
due to this inertial field, $\varphi({\bf r})= \int\limits_{{\bf
r}_0}^{{\bf r}}{\bf A}({\bf s})d{\bf s}$ which has the same origin
as the Aharonov-Bohm effect \cite{aharonov}. The inertial field
${\bf A}$ was found to be equal to the momentum ${\bf k}$ for the
free electron. A system of $N$ free electrons will be determined
in reciprocal space by a sphere of finite radius. The model in
this case includes the Pauli exclusion principle in a natural
manner. The general problem can be formulated in terms of the
local charge density distribution and the local complex vector
field. From the results for a homogeneous electron gas we derived
the solution for the vector field. This allowed to construct a
local density approximation for the general problem, which is
stated exclusively in terms of the density of charge, its
derivatives, and auxiliary quantities inferred from a homogeneous
electron gas.

These results are quite unexpected. Although it is probably too
early to appreciate the full extent of these findings, some general
points can be made.

The existence of inertial fields ${\bf A}$, related to electron
motion, seems to relate to the problem of finding transferable
kinetic energy functionals in DFT. If, as it seems, a crucial part
of the physical picture has so far been omitted, then it is not
surprising, that so far every formulation has been somewhat less
than transferable. It is also important to note that the central
result of the Hohenberg-Kohn theorem, i.e. the proof that the
groundstate charge density minimizes the total energy, remains
untouched by our findings. A complex phase added to the root of the
charge density will not alter the charge density at any point. In
this case the total energy of a given charge distribution is still a
minimum, even though additional contributions due to the complex
phase exist.

From a practical point of view it could substantially aid in the
development of local many-electron frameworks. The inertial fields
are related to electron motion; as they are fields, they should be
superimposable and have to comply with boundary conditions. Electron
charge by itself does not have the same constraints. The well known
and extensively developed framework of classical electrodynamics may
in this case also become useful for the study of field properties in
quantum mechanical systems. The field ${\bf A}$ is related to
magnetic properties of electrons, even if, as in case of free
electrons or hydrogen-like electrons, the corresponding magnetic
field ${\bf B}$ is zero. It may well be that the ultimate answer to
the question, what spin actually is, will come from a closer
analysis of the field under varying external conditions. In
addition, the finding could lay to rest a century-old problem, that
of the self-energy of electrons due to electrostatic interactions.
So far, all local electron models could not resolve the problem, why
electrons would retain their size. Here, we found that the effective
potential $v_{eff} = v_{gxc} + v_H = 0$ for single electrons. This
means, that no interactions exist between different regions of one
extended electron. In principle, this allows for any size of
electrons.

From a purely mathematical perspective, and in view of direct
methods in density functional theory, the proof of a successful
application to general condensed matter problems is still missing.
Given that the restrictions this relation imposes on the charge
density distribution are much more stringent than in current
models, and that any direct method is potentially much more
efficient than a method based on solutions of the Kohn-Sham
equations, the results seem promising. Progress, in this respect,
will depend on the development of transferable and precise
parameterizations of the general exchange-correlation potential,
or the vector field ${\bf A}$. A task, which has yet to be
accomplished.

From a physical and more fundamental perspective, the emerging
picture is quite intriguing. A quantum mechanical system, which
finds its groundstate by equilibrating energy density throughout,
is in principle not very different from a thermodynamic system of
molecules, which accomplishes exactly the same on the level of
molecular motion. There is thus no fundamental difference, if one
goes from the analysis of isolated molecules to the analysis of
electron charge. Furthermore, it is quite difficult to explain to
a layman, why an electron, as a point particle, does not
immediately attach itself to the nucleus of a hydrogen atom. The
present framework removes this problem, as well as the seeming
contradiction, that extended electrons, with their inherent
Coulomb repulsion, are nevertheless stable.

However, here we should add a note of caution. Even though the
results obtained so far look very promising, the success of the
framework will ultimately depend on a very pragmatic fact: whether
it can be used to make accurate predictions of solid state
properties, and whether these predictions are in line with
experimental data.\\

\section*{Acknowledgements}

WAH is indebted to George Darling, Jacob Gavartin, Andres Arnau and
Pedro Echenique for stimulating discussions, criticism, and support,
which to a large extent initiated or focussed research in this
particular direction. He is supported by a Royal Society University
Research Fellowship. KP is funded by the European Commission under
project NMP3-CT-2004-001561.

\end{document}